%% file: main.tex
\documentclass{article}

\usepackage{nac_preprint}
\usepackage[utf8]{inputenc} 
\usepackage[T1]{fontenc}    
\usepackage{hyperref}       
\usepackage{url}            
\usepackage{booktabs}       
\usepackage{amsfonts}       
\usepackage{nicefrac}       
\usepackage{microtype}      

\usepackage{comment}
\usepackage{tcolorbox}
\usepackage{booktabs} 
\usepackage{subcaption}
\usepackage{amsmath}
\usepackage{amssymb}
\usepackage{cleveref}
\usepackage{mathtools}
\usepackage[toc,page]{appendix}
\usepackage{enumitem}
\usepackage{graphics}
\usepackage{graphicx}
\usepackage{xcolor}
\usepackage[export]{adjustbox}
\usepackage{algorithm}
\usepackage{algorithmicx}
\usepackage[noend]{algpseudocode}
\usepackage{todonotes}
\algnewcommand{\LineComment}[1]{\State \(//\) #1}
\algnewcommand{\RLineComment}[1]{\State \(\triangleright\) #1}
\usepackage{bm}
\usepackage{multirow}

\usepackage{wrapfig}
\usepackage{ulem}
\usepackage{makecell}
\usepackage{authblk}

\newcommand{\customfootnotetext}[2]{{
  \renewcommand{\thefootnote}{#1}
  \footnotetext[0]{#2}}}

\usepackage{etoolbox}
\usepackage{tikz}
\usetikzlibrary{tikzmark}
\usetikzlibrary{calc}

\errorcontextlines\maxdimen

\newcommand{\ALGtikzmarkcolor}{black}
\newcommand{\ALGtikzmarkextraindent}{4pt}
\newcommand{\ALGtikzmarkverticaloffsetstart}{-.5ex}
\newcommand{\ALGtikzmarkverticaloffsetend}{-.5ex}
\makeatletter
\newcounter{ALG@tikzmark@tempcnta}

\newcommand\ALG@tikzmark@start{%
    \global\let\ALG@tikzmark@last\ALG@tikzmark@starttext%
    \expandafter\edef\csname ALG@tikzmark@\theALG@nested\endcsname{\theALG@tikzmark@tempcnta}%
    \tikzmark{ALG@tikzmark@start@\csname ALG@tikzmark@\theALG@nested\endcsname}%
    \addtocounter{ALG@tikzmark@tempcnta}{1}%
}

\def\ALG@tikzmark@starttext{start}
\newcommand\ALG@tikzmark@end{%
    \ifx\ALG@tikzmark@last\ALG@tikzmark@starttext
    \else
        \tikzmark{ALG@tikzmark@end@\csname ALG@tikzmark@\theALG@nested\endcsname}%
        \tikz[overlay,remember picture] \draw[\ALGtikzmarkcolor] let \p{S}=($(pic cs:ALG@tikzmark@start@\csname ALG@tikzmark@\theALG@nested\endcsname)+(\ALGtikzmarkextraindent,\ALGtikzmarkverticaloffsetstart)$), \p{E}=($(pic cs:ALG@tikzmark@end@\csname ALG@tikzmark@\theALG@nested\endcsname)+(\ALGtikzmarkextraindent,\ALGtikzmarkverticaloffsetend)$) in (\x{S},\y{S})--(\x{S},\y{E});%
    \fi
    \gdef\ALG@tikzmark@last{end}%
}

\apptocmd{\ALG@beginblock}{\ALG@tikzmark@start}{}{\errmessage{failed to patch}}
\pretocmd{\ALG@endblock}{\ALG@tikzmark@end}{}{\errmessage{failed to patch}}
\makeatother

\title{Predictive Coding with Spiking Neural\\ Networks: a Survey}
\author[1]{Antony W. N’dri}
\author[2]{William Gebhardt}
\author[1]{Céline Teulière}
\author[3]{Fleur Zeldenrust}
\author[4]{\\Rajesh P. N. Rao}
\author[1,5,\textdagger]{Jochen Triesch}
\author[2,\textdagger,*]{Alexander Ororbia}
\affil[1]{Universit\'{e} Clermont Auvergne, Clermont Auvergne INP, CNRS, Institut Pascal, F-63000 Clermont-Ferrand, France}
\affil[2]{Rochester Institute of Technology, Rochester, NY 14623, USA}
\affil[3]{Donders Institute for Brain, Behaviour and Cognition, Radboud University, Nijmegen, the
Netherlands}
\affil[4]{Paul G. Allen School of Computer Science and Engineering, University of Washington Seattle, Washington, USA}
\affil[5]{ Frankfurt Institute for Advanced Studies, Frankfurt am Main, Germany}
\affil[*]{Corresponding author: ago@cs.rit.edu}

\begin{document}

\setlength{\abovedisplayskip}{0.065cm}
\setlength{\belowdisplayskip}{0pt}

\maketitle

\customfootnotetext{\textdagger}{Authors contributed equally.}

\begin{abstract}
In this article, we review a class of neuro-mimetic computational models that we place under the label of \emph{spiking predictive coding}. Specifically, we review the general framework of predictive processing in the context of neurons that emit discrete action potentials, i.e., spikes. Theoretically, we structure our survey around how prediction errors are represented, which results in an organization of historical neuromorphic generalizations that is centered around three broad classes of approaches: prediction errors in explicit groups of error neurons, in membrane potentials, and implicit prediction error encoding. Furthermore, we examine some applications of spiking predictive coding that utilize more energy-efficient, edge-computing hardware platforms. Finally, we highlight important future directions and challenges in this emerging line of inquiry in brain-inspired computing.
Building on the prior results of work in computational cognitive neuroscience, machine intelligence, and neuromorphic engineering, we hope that this review of neuromorphic formulations and implementations of predictive coding will encourage and guide future research and development in this emerging research area.

\keywords{Predictive coding \and Spiking neural networks \and Efficient coding \and Neuromorphic computing \and Brain-inspired computing}
\end{abstract}



\section{Introduction}
\label{sec:intro}

In recent years, there has been a surge of interest in predictive coding and predictive processing approaches as theories of brain function \cite{huang_predictive_2011,spratling2017review,keller2018predictive,millidge2021predictive,jiang_predictive_2022,salvatori2023brain}. These theoretical ideas have spawned many experimental investigations into the question to understand if and how such concepts are implemented in brains. Unfortunately, the terms `predictive coding' and `predictive processing' are not used consistently in the literature and different researchers use them in different ways. While some authors, in particular in the psychological sciences, sometimes use such terms to refer rather broadly to the idea that brains make predictions, e.g., neural activity represent future rewards or the outcomes of an individual's actions, there are also authors, who use these terms for precisely defined algorithms and computational frameworks. One reason for the broad range of meanings of the term ``predictive coding', in particular, is its historical origins. Generally, there is a reasonable degree of consensus related to the notion that the brains of mammals (including humans) are capable of making and using predictions for generating adaptive behavior. A more specific definition of predictive coding is that of Rao and Ballard \cite{rao1999predictive}, who introduced a notion of predictive coding inspired by  efficient coding ideas \cite{Attneave1954,Barlow1961,srinivasan_predictive_1982}. Efficient coding theories postulate that brains try to save energy by exploiting redundancies in sensory information and recoding the sensory input in a more efficient manner (e.g., using less activity). Time-varying sensory inputs generally contain predictable structure due to the physics of the natural world and how we sense the world, leading to redundancies across time that can be exploited for a more efficient code. Rao and Ballard's idea was to implement a hierarchical generative model of the visual world in which Bayesian inference is performed using prediction errors \cite{rao-ballard-NC-1997,rao1999predictive,rao_optimal_1999}; as a result, predictable information is removed and only prediction errors are passed to higher levels of processing, thereby obtaining an efficient code. Friston's subsequent work \cite{friston2005energy,friston2008variational} extended Rao and Ballard's framework from maximum a posteriori (MAP) inference to full-fledged probabilistic inference via a particular variational approximation of hierarchical Bayesian inference.  These two lines of research are at the origin of much of the subsequent work in the ``predictive coding'' or ``predictive processing'' domain, but many variants and extensions have been formulated since then (see \cite{spratling2017review,jiang_predictive_2022,salvatori2023brain} for reviews of these). For the purpose of this survey, we regard the notion of learning from or with sensory \emph{prediction errors}, which drive the nature and form of the requisite neural message-passing structures, as the defining feature of the general notion of ``predictive coding.'' 

From the early days of the original formulation of predictive coding mentioned above to the current day, most work has considered implementations of the aforementioned ideas in connectionist neural networks composed of elementary processing units that permit continuous or graded activation levels. The activation level of such units is sometimes considered to represent the time-averaged firing rate of an individual neuron or a small population of neurons with similar tuning properties. Real brains work with spiking neurons, however, whose activity over short time scales of a few milliseconds is essentially binary --- either a spike is emitted or not. Information processing and learning in spiking neural networks are central topics in computational neuroscience and neuromorphic engineering research. A key motivation for studying spiking neural networks, in addition to their higher degree of biophysical realism as compared to connectionist networks, is that they may hold the key to understanding the brain's remarkable energy efficiency, although whether predictive coding actually results in a reduction of energy consumption is still a matter of debate (see \cite{kwisthoutComputationalResourceDemands2020a} for discussion). 


Given that at least part of the predictive coding literature is rooted in theories of energy efficient information coding and that information coding with spikes is thought to contribute to the brain's high degree of energy efficiency, it is quite natural to investigate spiking neural network formulations and implementations of predictive coding ideas. Surprisingly, however, this area has remained under-studied and the literature is fragmented into different, seemingly independently developing strands. Here, we aim to survey the existing approaches and organize them into different classes, discuss their respective advantages and shortcomings, and point out gaps in our current understanding. We hope that this will facilitate future research in this important and exciting area. 

\begin{figure}
\centering
\includegraphics[width=11cm]{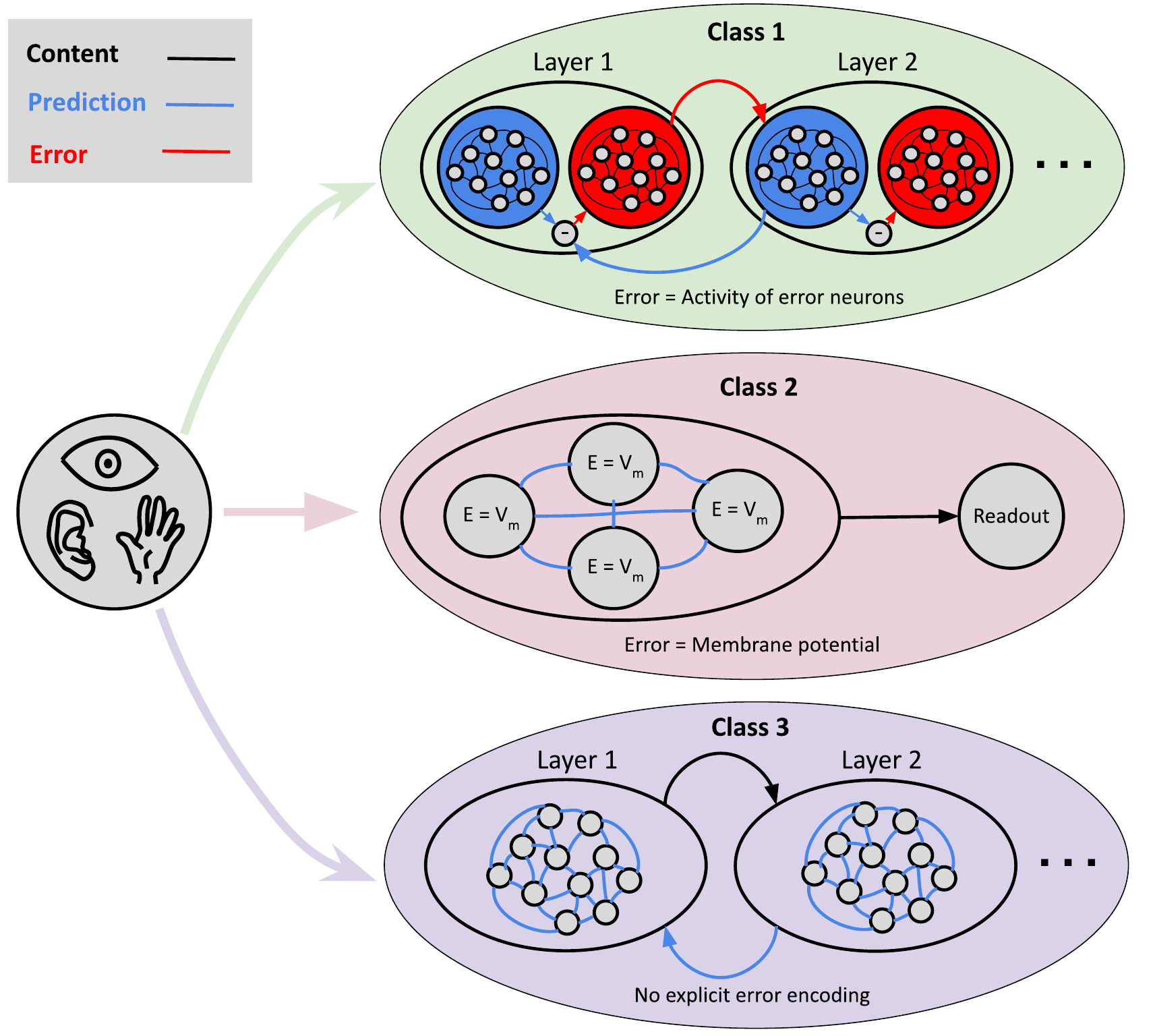}
\caption{
All models discussed in this survey deal with models that optimally or efficiently represent (functions of) stimuli. We organize the models discussed in this survey according to how the prediction error is formulated or represented. In Class 1, there are explicit groups of neurons coding for prediction errors. In Class 2, prediction errors are represented in the membrane potentials of individual neurons. In Class 3, prediction errors are not represented explicitly, but rather treated as error-like responses that are modelled in, for instance, competition between inhibitory and excitatory synapses.
 }
\label{fig:overview}
\vspace{-0.4cm}
\end{figure}

Since a common denominator of predictive coding approaches (according to our definition) is the notion of prediction error which organizes inference and learning, we structure our survey around how prediction errors are represented and used in different schemes (in 
\textbf{Box 1} 
we review and formally discuss rate-coded predictive coding). Specifically, we distinguish three broad classes of approaches. The first class uses explicit prediction error neurons, whose firing activity codes for the prediction errors. In the second class of approaches, prediction errors are not encoded by the spiking activity of neurons but by their membrane potentials. In the third class, we group alternative approaches, where prediction errors are encoded implicitly in the spiking neural network. We remark that a fourth class of networks are those that predict the future but do not effectively encode prediction errors. Such spiking networks \cite{bialekEfficientRepresentationDesign2006, heegerTheoryCorticalFunction2017, palmerPredictiveInformationSensory2015, salisburyOptimalPredictionRetina2016}
select for predictive features rather than surprising features. 
Depending on the level of noise, this can lead neurons that learn to either correlate or decorrelate their input (for a review see \cite{chalkUnifiedTheoryEfficient2017}). As these networks do not perform predictive coding, but rather engage in a form of `coding for the future', they will not be discussed here.

The remainder of this article is organized in the following manner. We start by reviewing some antecedents and early models of predictive coding with spiking neural networks in Sec.~\ref{sec:Historical}. Following this, in Sec.~\ref{sec:Class1} we present and study our first class of predictive coding approaches, which employ explicit error neurons. Section~\ref{sec:Class2} covers approaches from the second class, i.e., those that utilize neurons' membrane potentials to represent prediction errors. In Sec.~\ref{sec:Class3}, we discuss approaches from the third class, where prediction errors are encoded only implicitly. Fig.~\ref{fig:overview} summarizes how these three classes encode prediction errors. Section~\ref{sec:Applications} briefly presents some applications of spiking predictive coding approaches. Finally, in Sec.~\ref{sec:Conclusion}, we synthesize and conclude the survey, highlighting important future directions and challenges facing this emerging line of scientific research and engineering in the context of brain-inspired machine intelligence, computational cognitive neuroscience, and neuromorphic engineering.


\section{Historical Antecedents and Early Works}
\label{sec:Historical}

Historically, there have been different attempts at moving classical predictive coding approaches to the spiking domain. These start with the earliest works of Rao, Ballard and collaborators, and continue on to Denève’s formulation of Bayesian inference on spiking neurons as a form of predictive coding. A common factor behind some of those historical approaches is an attempt to explicitly define how spikes encode information, e.g. as probabilities. 

Predictive coding finds its roots in the idea of maximizing the probability of predicting the future inputs into a network. Similarly to this standard approach, early spiking predictive coding models are based on the idea that a neuron spikes to improve predictions of its future inputs. In addition to those early spiking predictive coding models, there has also been general prior work on constructing spiking neural networks that learn efficient sensory codes. Some of those efforts were aimed at compressing information, such that neurons respond only to the non-redundant (and therefore unpredictable, as in standard predictive coding) parts of stimuli. Such efficient spiking models inspired recent spiking predictive coding models -- notably those of Class 3, as per our survey's organization -- that deviated from the classical stance of predictive coding with explicit, separate prediction and error neurons. 

\begin{figure}
\includegraphics[width=\textwidth]{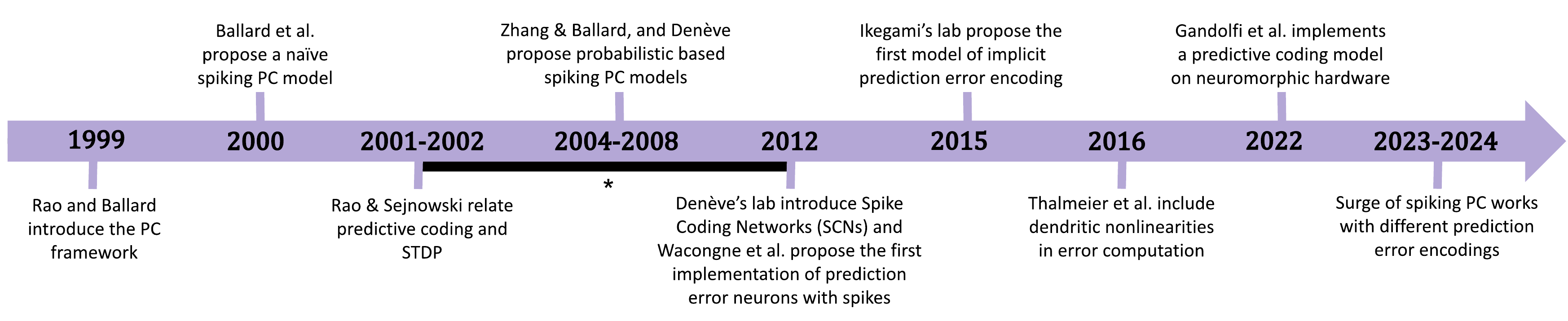}
\caption{{\bf History of spiking predictive coding models.} Timeline of key research efforts related to the  development of spiking predictive coding.  The black bar extending from 2001 to 2012 marked with an asterisk (*) indicates early works on sparse codes, ICA and PCA with spiking neurons by various groups.
 }
\label{fig:timeline}
\vspace{-0.4cm}
\end{figure}

In this work, we separate the aforementioned research efforts into the following three historical groupings: 
\textbf{1)} early work on learning efficient sensory codes, 
\textbf{2)} early work by Ballard, Rao and co-workers, and 
\textbf{3)} early work by Denève. We first review below the different models that these efforts encompass and then discuss how they relate to recent work. In Fig.~\ref{fig:timeline}, key efforts in this body of early work are presented historically (particularly in the first half of the timeline), contextualizing the works that came later (which are presented in the second half of the timeline).

\paragraph{Early work on learning efficient sensory codes in spiking neural networks.}
As discussed above, predictive coding approaches are, at least in part, rooted in more general efficient coding ideas. In this work, we treat the phrase ``efficient coding'' in terms of the efficient coding hypothesis, where sensory neurons are viewed as encoding maximal information about sensory inputs given (internal) constraints such as only using few spikes. 
Work on unsupervised learning of efficient sensory codes in spiking neural networks can therefore be seen as a precursor to full-fledged predictive coding approaches using spiking neuronal units. Indeed, there exists a body of relevant work implementing (approximations of) various unsupervised learning approaches such as principal component analysis (PCA) \cite{hotelling1933analysis}, sparse coding (SC) and dictionary learning \cite{olshausen2004sparse,lee2006efficient}, and independent component analysis (ICA) \cite{bell1997independent,hyvarinen2009independent} in spiking neural networks. All these approaches somehow try to convey a maximum amount of information about a stimulus while minimizing the resources necessary to do so. In this sense they can be considered {\em efficient}.

Toyoizumi et al.\ derived a learning rule for a stochastic spiking neuron that maximizes the mutual information between its input and output spike trains while maintaining its firing rate within a desired regime \cite{toyoizumi2005generalized}. The group led by Maass extended this approach so as to allow a neuron to maximize the mutual information between its input and output spike trains while simultaneously minimizing its output's mutual information with other spike trains or other time-varying signals, ultimately connecting the approach to the information bottleneck method \cite{tishby2000information} and ICA \cite{klampfl2006information,klampfl2009spiking}. They also formally derived a simplified linearized version of the model establishing a link to PCA \cite{buesing2007simplified} (by deriving an update rule based on a PCA objective and showing that this rule is equivalent to the PCA-oriented covariance
rule for non-spiking neurons \cite{sejnowski1989hebb}). Several other works have implemented forms of SC and ICA in spiking neural networks. Perrinet and colleagues \cite{perrinet2002sparse,perrinet2004sparse} used a code based on spike latencies, i.e, a rank order coding scheme,  in combination with a greedy form of matching pursuit \cite{mallat1993matching} in order to learn visual receptive fields; a similar approach was used for auditory signals \cite{Smith2005}. Matching pursuit sparsely activates only units that encode information that has not been encoded yet. In this sense it leads to an efficient and non-redundant code.
In a complementary line of work, Zylberberg et al.\ presented a spiking neural model that evolved according to local learning rules, reproducing many features of so-called simple cells in the mammalian visual cortex when trained on natural images \cite{zylberberg2011sparse}. Hu et al.\ presented a hybrid distributed algorithm (HDA), which learned an efficient sensory code and whose operation is equivalent to a network of integrate-and-fire neurons \cite{hu2012network}. Finally, Savin et al.\ demonstrated an approximation of ICA with a network of spiking neurons that learned by combining local and biologically-plausible rules including spike timing-dependent plasticity (STDP), synaptic normalization, and plasticity of intrinsic excitability with an adaptive form of lateral inhibition \cite{savin2010independent}. Similar learning systems, based on only local plasticity rules, were crafted by Clopath et al., who introduced a voltage-dependent STDP rule \cite{clopath2010connectivity}. Subsequently, Gilson et al.\ formally related STDP to both PCA and ICA \cite{gilson2012spectral}. These and other related efforts set the stage for research on spiking neural networks that attempted to implement full-blown predictive coding schemes.


\paragraph{Early work by Ballard, Rao, and collaborators.} While the predictive coding framework was formally introduced as a rate coded method by Rao and Ballard in their seminal paper in 1999 \cite{rao1999predictive}, research has provided growing evidence of the existence of precise spike timings in the neural code \cite{mainen1995reliability, cox2000action}. Such evidence motivated early work on spiking predictive coding models. Ballard et al. proposed in \cite{ballard2000} one of the first models in which an input is encoded as synchronous codes that are repeated in time. In this model, spiking neurons are represented by binary units whose values and synaptic weights are learnt through error minimization based on the minimum description length principle. This work entailed the learning of a network that could cancel sequential spikes from binary encoded images. Following this effort, still with the same motivations, Zhang and Ballard proposed another spiking predictive coding model \cite{zhang2004}. They suggested a functional interpretation of spikes, by which neurons' spikes explicitly encode information as probabilities. 
They learnt synaptic weights 
by performing gradient descent in a recurrent network modeling the lateral geniculate nucleus (LGN) and the primary visual cortex (V1). Their model could account for a number of biological observations such as simple cell-like receptive fields and neuronal activities that exhibited oscillations in the gamma frequency band. 
In parallel, Rao and Sejnowski, in a series of papers, established a link between spike timing-dependent plasticity (STDP) and prediction-centered learning. They employed a temporal difference (TD) learning rule in tandem with compartmental neuron models demonstrating that this could be equivalent to STDP by replicating pre/post spike pairings \cite{rao1999predictivespike, rao2001spike}. The authors highlighted the property of STDP rules in learning predictions and suggested that STDP-induced inhibition and excitation could be akin to positive and negative feedback observed in standard predictive coding. Later work considered a pool of spiking neurons driven by moving visual patterns \cite{rao2002}. Their results showed that neurons spiked earlier and that feed-forward input tended to be suppressed, as prescribed by standard predictive coding. Interestingly, the STDP-learnt connections were direction selective and consistent with monkey visual cortex findings. 

While recent models use Hebbian learning, e.g. STDP, as their primary learning rule instead of gradient descent, as in Ballard's initial approaches, the information-theoretic view proposed in these early models, including those based on STDP-induced prediction learning, are still actively used (see models of class 1 and 3). 

\begin{figure}[!t]
\begin{tcolorbox}[width=\linewidth, sharp corners=all,left=0pt,right=0pt,boxrule=0pt, colback=white!95!black]
\textbf{Box 1: The Dynamics of Mechanistic Predictive Coding.} Formally, the general mathematical formulation of mechanistic predictive coding, from the perspective of efficient coding as developed by Rao and Ballard, typically starts from the specification of a variational free energy functional (see \cite{rao1999predictive,bogacz2017tutorial,salvatori2023brain}, for details) such as the following:
\begin{align}
\mathcal{F}(\Theta) = \sum^L_{\ell=0} \frac{1}{2\Sigma^\ell} \sum^{\mathcal{J}_\ell}_{i=1} \big( \mathbf{z}^\ell_i(t) -  \mathbf{\bar{z}}^\ell_i \big)^2\label{eqn:vfe_functional}
\end{align}
where $l$ denotes layers of the model, $i$ indexes the elements of a vector , $\mathbf{\bar{z}}^\ell = \mathbf{W}^\ell(t) \cdot \phi^{\ell+1}(\mathbf{z}^{\ell+1})$ are the predictions (emitted by layer $\ell+1$) of layer $\ell$'s activity values, $\mathbf{z}^{0}$ corresponds to the model's input, and $\Theta = \{\mathbf{W}^\ell(t)\}^L_{\ell=1}$ is the set of all of the connection weight matrices of the model -- $\mathbf{W}^\ell(t)$ contains the generative/predictive synapses. Here, we use $\cdot$ to denote the multiplication of matrices/vectors (below, we use  
$\odot$ to represent the Hadamard product, and the transpose of a vector $\mathbf{v}$ is represented as $(\mathbf{v})^{\mathsf{T}}$). 
Equation \ref{eqn:vfe_functional} effectively states that the sum of weighted mean squared errors -- the differences between each neural activity $\mathbf{z}^\ell_i(t)$ and a prediction of it (emitted by another layer of neurons) -- at each layer of the predictive coding circuit contributes locally to the neural model's global energy value at any time step $t$. 
In effect, this functional assumes that the predictive coding circuit embodies/learns a hierarchical multivariate Gaussian generative model. Note that this is one of the simplest forms that the energy functional can take; other ones are possible depending on the choice of distributions assumed at each layer. 
In the predictive coding literature, it is common to assign the inner vector subtraction of Equation \ref{eqn:vfe_functional} to a multivariate variable, i.e., $\mathbf{e}^\ell = \mathbf{z}^\ell(t) - \mathbf{\bar{z}}^\ell$; this multivariate variable is represented by  the \emph{error neurons} and the first derivative of Equation \ref{eqn:vfe_functional} $\frac{\partial \mathcal{F}(\Theta)}{\partial \mathbf{\bar{z}}^\ell} = \frac{1}{\Sigma^\ell}\Big(\mathbf{z}^\ell(t) - \mathbf{\bar{z}}^\ell\Big)$ is referred to as the precision-weighted error. We point out that this basic treatment of predictive coding assumes a fixed scalar precision $\frac{1}{\Sigma^\ell}$; however, this need not be the case and $\Sigma^\ell$ could instead take the form of a learnable synaptic (covariance) matrix, which is done in some variants of predictive coding \cite{rao-ballard-NC-1997,ororbia2022ngc,salvatori2023brain}.

In service of optimizing the above free energy functional, the neurons in any layer of a predictive coding circuit evolve according to the following time-varying dynamics \cite{rao1999predictive} (which also models a form of short-term plasticity):
\begin{align}
    \tau_m\frac{\partial \mathbf{z}^\ell(t)}{\partial t} = -\gamma \mathbf{z}^\ell(t) + \mathbf{d}^\ell \odot \partial \phi^\ell(\mathbf{z}^\ell(t))  -\mathbf{e}^\ell(t)  \label{eqn:pc_state_update}
\end{align}
where $\tau_m$ is the cellular membrane time constant and $\mathbf{d}^\ell = \Big(\mathbf{W}^{\ell}(t)\Big)^{\mathsf{T}} \cdot \mathbf{e}^{\ell-1}(t)$ contains the perturbations produced by a projection of the error unit values at layer $\ell-1$; notice that the error mismatch values are passed backwards along a local feedback pathway made up the transpose of the same generative/forward synapses. Inference and learning in the predictive coding model above proceeds in a (dynamic) expectation-maximization manner \cite{dempster1977maximum,friston2008variational}. When processing sensory data, the above Equation \ref{eqn:pc_state_update} is run repeatedly and iteratively via Euler integration, for all layers of neurons in parallel, over a window of time of length $T$ -- the stimulus presentation time -- before computing the updates to synaptic connection strengths. Specifically, the forward generative synapses are adjusted according to a simple two-factor Hebbian rule \cite{rao1999predictive}:
\begin{align}
    \tau_w \frac{\partial \mathbf{W}^\ell(t)}{\partial t} &= -\gamma_w \mathbf{W}^\ell(t) + \mathbf{e}^{\ell}(t) \cdot \Big(\mathbf{z}^{\ell-1}(t)\Big)^{\mathsf{T}}  \label{eqn:pc_gen_synaptic_update} 
\end{align}
where $\tau_w$ is  the generative  synaptic plasticity time constant and $\gamma_w$ controls the degree to which decay is applied to the generative synaptic update. In essence, the above equations indicate that plasticity for a synapse is based on error neuron activity and pre-synaptic state unit values. 
\end{tcolorbox}
\vspace{-0.5cm}
\end{figure}

\paragraph{Early work by Denève.}\label{par:BN}
Crucially, Denève proposed, in a series of papers \cite{deneve2004, deneve2008a, deneve2008b}, an extension of Bayesian inference in spiking neurons to predictive coding. Bayesian inference consists of making statistical inferences that are based on the confidence that a particular hidden cause is present or a response or event will happen according to previous beliefs about the world, i.e. priors. Denève’s model argues that a neuron should only spike if that spike conveys new information about the presence of hidden causes. It is efficient as it should encode only information that could not be explained by previous spikes.

Denève formulates the Bayesian model as follows: the neuron tracks both the log-odds ratio of a hidden cause given all past synaptic inputs and the log-odds ratio of the same hidden cause based on its own output spike train.  When the log-odds ratio based on the input is higher than that based on the output plus a threshold, the neuron spikes. The `membrane potential' is interpreted as the difference between the two log-odds ratios. The predictive coding component is effectively in this `membrane potential': the comparison of the estimate of the hidden cause based on the input and that based on the neuron's own output. The estimate of the hidden cause based on the output can be interpreted as a moving threshold or an increasing refractory period. Hence, a spike is fired only when a neuron encounters input that it has not conveyed yet in its output. The weights of the inputs to the Bayesian neuron can be learnt without supervision, using for example an expectation maximization (EM) algorithm, as their updates only depend on the statistics of the input spike train. This approach to Bayesian inference thus maximizes the probability of explaining hidden causes from a neuron’s spike train and is efficient, as the neuron responds only to stimuli it did not convey yet. Notably, the principles behind the Bayesian neuron have been implemented in an experimental setup \cite{Zeldenrust2017, dasilvalantyerDatabankIntracellularElectrophysiological2018, zeldenrustTuningTuningHow2024}, where it was shown that the aforementioned forms of adaptation indeed occur in cortical neurons and increase the efficiency of information processing.

As alluded to in the previous paragraph, Denève relates the spiking dynamics of the Bayesian neuron model to leaky integrate-and-fire neurons by showing that the membrane potential of those neurons can be interpreted as the difference between the log-odds ratio of a stimulus being present based on the input (synaptic evidence accumulation) and the log-odds ratio based on the neuron's own prediction of the presence of the stimulus. This introduces the formalism of the membrane voltage as a prediction error used in some of the recent work in spiking predictive coding (see Class 2). 

\paragraph{Conclusion} A wide variety of works have attempted to move predictive coding to spiking networks. 
It is important not to mistake general Bayesian models of inference using spiking implementations (e.g. \cite{rao2004hierarchical}) for predictive coding models, as most of those general Bayesian models explicitly decode information from spiking networks without any constraints on efficiency, contrary to predictive coding models. While the research efforts described were introduced around the same time frame as classical predictive coding (see timeline in Fig. \ref{fig:timeline}), they motivated only a few researchers at that time to explore spiking networks. As there is now more interest in spiking neural networks and predictive coding, it is important to revisit the historical efforts described above in this section and observe their relationship with present spiking models of predictive coding. For instance, the work of Denève gave rise to Class 2, composed of so-called spike coding networks, which we review in Sec.~\ref{sec:Class2}. Similarly, the different works related to efficient sensory codes as well as the seminal works of Ballard and co-workers are precursors to the models of Class 1 and Class 3, which we review in Sec.~\ref{sec:Class1} and \ref{sec:Class3}. Overall, these early works exposed the difficulties of applying predictive coding to spiking networks and introduced many of the ideas that underpin more recent forms of spiking predictive coding.

\section{Spiking Predictive Coding with Explicit Error Neurons (Class 1)}
\label{sec:Class1}


\begin{table}[!t]
  \footnotesize
  \centering
  \resizebox{\columnwidth}{!}{%
  \begin{tabular}{cccccc}
	\toprule
	Authors  & Prediction & Neuron type & Dale’s law?  & Learning rule & Explained phenomena/Application\\
	\midrule
	Wacongne et al. \cite{wacongne2012neuronal, wacongne2016predictive} & Top-down input & Izhikevich & yes & Fixed weights and STDP & \thead{How to apply predictive coding to\\ reproduce mismatch negativity and \\how it relates to schizophrenia}\\
    \midrule
    Lan et al. \cite{lan2022} & \thead{Top-down input\\(only during learning)} & IF & yes & Free energy principle & \thead{How to replace backpropagation \\with predictive coding}\\
    \midrule
    Fraile et al. \cite{fraile2023competition} & Lateral input & LIF & yes & Fixed weights & \thead{Observation of explicit prediction error-like \\neurons in a bio-plausible V1 network}\\
    \midrule
    Lee et al. \cite{lee2024predictive} & Top-down input & ADEX & yes & Rate-based hebbian learning & \thead{Generation/ reconstruction of \\visual input samples}\\
    \midrule
    Ororbia \cite{ororbia2023spiking} & Top-down input & LIF & yes & ST-LRA (event-driven local rule)& \thead{Generation/ reconstruction of visual\\ input samples, continual classification}\\
	\bottomrule 
  \end{tabular}
  }
  \vspace{0.1cm}
  \caption{Models using explicit error neurons (Class 1)}
  \vspace{-0.5cm}
\end{table}

The first class of spiking predictive coding models that we will consider are those in which prediction errors are coded by designated error neurons. In this category, error neurons are explicitly defined and charged with computing the mismatch between the sensory representation of an external input and predictions (produced by top-down pathways or feedback) of this sensory representation. Standard predictive coding appeared difficult to reconcile with spiking networks, since it requires error neurons that can encode both positive and negative prediction errors. How can the activity of spiking error neurons, whose firing rates can only be positive, encode both positive and negative numbers? One initial solution might be to represent the signed error as the positive or negative deviation of the firing rate with respect to a baseline firing rate value (that represents zero or the absence of prediction error). 
However, to solve this problem, some models of this first class define two types of error neurons for representing positive 
vs.\ negative prediction errors (inspired by on-center off-surround and off-center on-surround types of cells in the retina), while others use rate-based error units in conjunction with spiking prediction units. The presence of such units in the cortex has been reported in certain studies but is still a highly debated subject in the field \cite{keller2018predictive}. We review below predictive coding models that try to apply this particular way of modeling error neurons in spiking systems.


\paragraph{Mismatch Negativity with Spiking PC.}
Wacongne et al. proposed the first spiking implementation of prediction error neurons \cite{wacongne2012neuronal}. Specifically, this mechanism mimicked so-called mismatch negativity (MMN) exhibited in a network of Izhikevich neurons. 
The authors considered a 3-layered network consisting of a prediction error layer, a prediction layer and a memory trace layer. In addition, a thalamic-like input was used to excite the network. In this architecture, prediction error neurons received feed-forward AMPA excitation from two populations of thalamic neurons as well as recurrent feedback GABA inhibition from prediction neurons. Prediction error neurons would then send back errors to prediction neurons which would subsequently update memory neurons. The memory neurons then accumulated information and sent this back via a recurrent excitatory pathway to the prediction neurons. In addition to those AMPA/GABA currents, the authors also consider an NMDA receptor-dependent current that is added when receiving excitation. While most of the model's connection weights are random, the synaptic connections between memory neurons and prediction neurons were adjusted with STDP. 
Importantly, this work was able to successfully replicate biological observations of MMN. Furthermore, while MMN could be also explained by synaptic habituation, this effort opposed this explanation of MMN and instead appealed to a fully predictive coding perspective. An MEG experiment was performed to show that a predictive model accounted for biological observations of MMN which could not be explained by synaptic habituation alone. Wacongne later also showed that their model could account for changes in MMN observed in people affected by schizophrenia \cite{wacongne2016predictive}.

\paragraph{Predictive Coding Error Neurons in a Bio-plausible V1 Model.} After the introduction of a biologically plausible model of V1 composed of leaky integrate-and-fire neurons by Billeh et al. in \cite{billeh2020systematic}, Fraile et al. studied this network to see if it could replicate observations of prediction error neurons and evaluated the conditions under which they emerge \cite{fraile2023competition}. In V1, such neurons have been reported in the supragranular layers (L2/3) during sensorimotor experiences \cite{jordan2020opposing}. Neurons in L2/3 receive top-down projections which makes them prime candidates for prediction error neurons \cite{rao1999predictive,jiang_predictive_2022}. Furthermore, L2/3 neurons have also been reported to be depolarized/hyperpolarized even during the absence of top-down motor input \cite{jordan2020opposing, muzzu2021feature}. 

Fraile et al.\ stimulated the network with different drifting gratings and observed the presence of two categories of neurons in L2/3: depolarized (the positive error neuron) and hyperpolarized (the negative error neuron). They verified excitatory/inhibitory connectivities and the synaptic strength of the neurons, determining that there were differences between the two categories which allowed them to act as error neurons even during bottom-up excitation alone. For instance, positive error/depolarized neurons had a strong connectivity to the internal granular layer (L4) and LGN, while negative error/hyperpolarized neurons were weakly connected in comparison. The authors also examined the connectivities and response dynamics of internal pyramidal cell layer (L5/6) neurons. Their results suggested that those neurons did not have the properties required for the emergence of error neurons. While the model used was biologically plausible, it only considered point neurons and not compartmental neuron models with dendrites, which have also been hypothesized to play an important role in the emergence of prediction error neurons.

\begin{figure}[!t]
\begin{tcolorbox}[width=\linewidth, sharp corners=all,left=0pt,right=0pt,boxrule=0pt, colback=white!95!black]
\textbf{Box 2: Spiking Neural Coding Network Dynamics and Hebbian Plasticity.} The spiking neural coding network centers around the notion of explicit error neurons, specifically those cast in terms of the following subtractive mismatch calculation:
\begin{align}
    \mathbf{e}^\ell(t) = -\big(\mathbf{z}^\ell(t) - \mu^\ell(t)\big) \label{eqn:spncn_error}
\end{align}
where the above is the fixed-point equation for the error neuron activity values and $\mu^\ell(t)$. The simplest form that $\mu^\ell(t)$ can take is a linear transformation, i.e., $\mu^\ell(t) = \mathbf{W}^T \cdot \mathbf{s}^{\ell+1}(t)$ which contains the predictions emitted by another layer (such as layer $\ell+1$) attempting to guess the spike trace of the neurons in layer $\ell$. 

Although the framework was agnostic to the type of spiking neuron model employed, it was originally formulated as a network of synaptic conductance-driven leaky integrators modeled in terms of the following two differential equations:
\begin{align}
    \tau_j \frac{\partial \mathbf{j}^\ell(t)}{\partial t} &= -\mathbf{j}^\ell(t) + f(\mathbf{e}^{\ell-1}(t); \Theta) \\
    \tau_m \frac{\partial \mathbf{v}^\ell(t)}{\partial t} &= -\mathbf{v}^\ell(t) + \Big( \mathbf{j}^\ell(t) R_m \Big) 
\end{align}
where the membrane potential $\mathbf{v}^\ell(t)$, driven by electrical current $\mathbf{j}^\ell(t)$, ultimately feeds into any SNM or spike emission function to produce a discrete action potential for time $t + \Delta t$, e.g., $\mathbf{s}^\ell(t + \Delta t) = \mathbf{v}^\ell(t) > \mathbf{v}_{\text{thr}}$ ($\mathbf{v}_{\text{thr}}$ is a vector containing threshold values, one per neuronal unit). Notice that the electrical current, which evolves with time, depends on a function $f$ of a previous layer $\ell-1$ of neuronal error activities $\mathbf{e}^{\ell-1}(t)$ (this function can be designed to be as simple as a linear transformation or a complicated set of nonlinear dendritic tree operations). Given a spike neuronal model, e.g., $\mathbf{s}^\ell(t + \Delta t) = \mathbf{v}^\ell(t + \Delta t) > \mathbf{v}_{\text{thr}}$ and a hyperpolarization step, such as $\mathbf{v}^\ell(t + \Delta t) = \mathbf{v}^\ell(t + \Delta t) \odot \big(1 - \mathbf{s}^\ell(t + \Delta t)\big)$, a (piecewise) variable trace of the spikes is then tracked according to the following:
\begin{align}
    \mathbf{z}^\ell(t + \Delta t) = \Big( \mathbf{z}^\ell(t) -\frac{\Delta t}{\tau_{tr}}\mathbf{z}^\ell(t) \Big) \odot \Big(1 - \mathbf{s}^\ell(t + \Delta t)\Big) + \mathbf{s}^\ell(t + \Delta t)
\end{align}
where $\tau_{tr}$ is the trace time constant (on the order of milliseconds) and  $\Delta t$ is the integration time constant. Given the trace above, error neuronal units are simply computed by comparing this trace with a set of prediction values $\mu^\ell(t)$ as per Equation \ref{eqn:spncn_error}. 

Once the error neural activities have been updated at any given step in time, synaptic efficacies may be adjusted with simple, local Hebbian rules that take the form:
\begin{align}
    \frac{\partial w_{ij}}{\partial t} = -\gamma w_{ij} + e_i s_j
\end{align}
noting that $\gamma$ is the synaptic weight decay coefficient and that two-term Hebbian product is between the error at post-synaptic neuron $i$ and pre-synaptic spike $j$.
\end{tcolorbox}
\vspace{-0.5cm}
\end{figure}

\paragraph{SpNCN: Spiking Neural Coding Network.} The computational framework proposed by Ororbia \cite{ororbia2023spiking} attempts to faithfully embody the core mechanisms of classical PC circuitry while casting them in the context of spike trains. Specifically, the spiking neural coding network (SpNCN) is a generalization of neural generative coding (NGC) \cite{ororbia2022ngc} in terms of spike neuronal models (SNMs), which are engineered mathematical functionals that elicit particular properties of spiking cells known in the computational neuroscience literature, e.g., leaky integrate-and-fire, spike-response kernels, etc. For each layer of SNMs within an SpNCN model, a local prediction is emitted as an electrical pulse to either directly-wired error neurons or to another set of predictive SNMs. Error neurons in the framework of the SpNCN notably operated on bio-chemical neuromodulatory/glutamate concentrations mathematically manifested as simple (variable) traces. Specifically, in an SpNCN, the traces were specifically designed to track post-synaptic spike activities, implying that neurons in one layer of an SpNCN were attempting to predict the spike timings of another layer. Desirably, the error signals produced at the trace level could also be leveraged in local, event-driven synaptic adjustment rules that were triggered at each simulation time step. In \textbf{Box 2}, 
we formally depict the core neuronal dynamics and local plasticity that mathematically characterized the SpNCN. Ororbia 
\cite{ororbia2023spiking}  demonstrated that such SpNCNs yielded classification performance that was competitive with deep networks and could further be modified to conduct a form of online continual adaptation when processing complex sensory input streams. 

\paragraph{SNN-PC: Spiking Neural Network for Predictive Coding.}
The spiking neural network for predictive coding (SNN-PC) model was proposed by Lee et al. \cite{lee2024predictive} as an advancement in biological realism. Through the use of a neuronal architecture built with adaptive exponential integrate and fire (AdEx) neurons, an NMDA-mediated component for synaptic transmission, and error units, the authors built a fully spiking predictive coding system.

One of the authors’ main contributions to improving biological realism comes from their reformulation of the traditional error unit. Specifically they separated every error unit into two subcomponents; one coding for positive error values and another for negative values. The two components work off of the same signals, bottom-up excitatory inputs and top-down inhibitory inputs, but the formulated combination of these inputs is inverted.

The other contribution of Lee et al. is their use of feedforward gist pathways in their model. It is believed that visual cortical processing takes place within two processes: a fast approximation of the world/niche and a slower, more detailed process. The second of these is tackled by the predictive coding model and the error nodes described previously. The first, higher level representation is crafted by the gist pathway. In the authors’ work the input image is transformed through sparse random connections, the gist unit, and then passed through to each of the representation units to help bias the representations towards a specific image.

\paragraph{PC-SNN: Predictive Coding in Spiking Neural Networks.}
In their recent work Lan et al. proposed an adaptation to the predictive coding algorithm designed for ANNs called the ``PC-SNN’’ \cite{lan2022}. PC-SNN uses a Hebbian-based rule to control its synaptic plasticity, which is influenced solely by local nodes. Specifically, the error node of the postsynaptic layer and the integrate-and-fire (IF) neuron of the previous layer are used to adjust the synaptic plasticity. In the PC-SNN model, the error nodes are a simple subtraction between the first spike timing of a postsynaptic neuron and the actual spike timing of that neuron.

Lan et al. focused on the biological concept of free energy proposed by Friston \cite{friston2005energy} to derive their learning rule. They defined the free energy of their problem to be the summation of errors both across all error nodes and across all time. They logically reason that the system is too complex to be able to jump to a steady state, but they can slowly adjust the synaptic efficacy to walk the model down towards the steady state; note that this objective is similar to that of the earlier SpNCN \cite{ororbia2023spiking} which also iteratively optimizes a temporal energy functional. Notably, the authors added two additional constraints to their model. The first constraint was that every IF neuron could only emit at most one spike for any given image input; this meant that, in order to perform classification, they used a first-to-spike (winner-take-all; where only one neuron can emit a spike at a time) coding scheme. The other constraint was applied to the synaptic updates -- the update to a synapse was masked so as to not produce a weight change if the predicted neuron spiked before the predicted timing.

\paragraph{Conclusion.} `Class 1' predictive coding networks represent some of the first efforts to cast prediction error (or free energy) minimizing dynamics in terms of systems of discrete, pulse-emitting neuronal systems. Ultimately, these models rest on the assumption that mismatch/error signals come from explicit error neurons (the sum of whose activities over time approximate a free energy function) and, as a result, build their message-passing architectures around this building block. The primary difference in these models is the tactic that they take in modeling these error neurons (for instance, some model these changes being produced by different concentrations of neurotransmitters through variable traces while others design units that encode differently the signs of the various error signals) and the biophysical mechanisms that they instantiate for synaptic plasticity as well as the transmission of feedback and expectation signals across the full neuronal structure. 

\section{The Membrane Potential as a Prediction Error (Class 2)}
\label{sec:Class2}

\begin{table}[!t]
  \label{tab:modelspotential}
  \footnotesize
  \centering
  \resizebox{\columnwidth}{!}{%
  \begin{tabular}{cccccc}
	\toprule
	Authors & Prediction & Neuron type & Dale’s law?  & Learning rule & Explained phenomena/Application\\
	\midrule
	  \thead{Boerlin et al. \cite{boerlin2011spike, boerlin2013}\\
    Calaim et al. \cite{calaimGeometryRobustnessSpiking2022}\\
    Barrett et al. \cite{barrettOptimalCompensationNeuron2014}} & Lateral input & LIF & No & Fixed weights & \thead{Efficient coding, irregular firing, \\trial-to-trial variability, robustness}\\
    \midrule
    \thead{Nardin et al. \cite{10_24072_pcjournal_69}\\
    Slijkhuis et al. \cite{slijkhuisClosedFormControlSpike2023}}  & Lateral input  & LIF & No & Fixed weights & \thead{How to apply SCNs to perform \\ polynomial dynamical systems or control}\\
\midrule 
   \thead{Bourdoukan et al. \cite{Bourdoukan2012}\\
   Alemi et al. \cite{alemiLearningArbitraryDynamics2017}\\
   Brendel et al. \cite{brendel2020learning}} & Lateral input & LIF & No &  Hebbian & \thead{Derive Hebbian learning rules for SCNs}\\
    \midrule
    Zeldenrust et al. \cite{zeldenrust2021} & Lateral input & Filter and fire & No & Fixed weights & \thead{Role of neural heterogeneity on efficiency and robustness \\ relation trial-to-trial variability and efficiency}\\
    \midrule
    Schwemmer et al. \cite{Schwemmer2014} &  Lateral input  & Conductance based & No & Fixed weights & \thead{How SCNs can be formulated \\ in biophysically plausible networks}\\
    \midrule
    Gutierrez and Denève \cite{gutierrezPopulationAdaptationEfficient2019} & Lateral input  & \thead{LIF + \\spike frequency adaptation} & No & Fixed weights & \thead{How spike-frequency adaptation can be understood \\ as a solution to the cost-accuracy trade-off}\\
     \midrule
    Koren and Denève \cite{koren2017computational} & Lateral input  & LIF &  No & Fixed weights & \thead{How delays can be compensated by spike cost }\\
    \midrule
    Chalk et al. \cite{Chalk2016} & Lateral input  & LIF &  Yes &  Fixed weights & \thead{Gamma oscillations as a hallmark of efficiency }\\
    \midrule
     Koren and Panzeri \cite{koren2022biologically} & Lateral input  & generalized LIF & Yes &  Fixed weights & \thead{How to perform general linear transformations}\\
    \midrule
    Rullán Buxó and Pillow \cite{rullanbuxoPoissonBalancedSpiking2020} & Lateral input  & Conditional Poisson neuron & No & Fixed weights & \thead{How to implement delay and \\positive correlations between similarly tuned neurons}\\
    \midrule
     Timcheck et al. \cite{ timcheck2022optimal} & Lateral input  & LIF & No & Fixed weights & \thead{Relation coding fidelity, noise and delay\\ there is an optimal noise level given a synaptic delay}\\
    \midrule
Thalmeier et al. \cite{thalmeierLearningUniversalComputations2016} & Lateral input  & LIF with non-linear dendrites& Yes saturating  & Hebbian &\thead{How to approximate non-linear systems} \\
    \midrule
     Mikulasch et al. \cite{mikulasch2023, mikulaschLocalDendriticBalance2021} & Apical dendritic input & LIF with dendrites & No & Hebbian & \thead{Error-based learning based on \\a dendritic balance of excitation and inhibition }\\
        \midrule
     Zhang and Bohte \cite{zhang2024energy} & Apical dendritic input &  \thead{LIF with dendrites + \\threshold adaptation} &  No & \thead{Loss minimization with \\forward propagation through time}& \thead{Predictive coding as a result of energy optimization}\\
	\bottomrule 
  \end{tabular}
  }
  \vspace{0.1cm}
  \caption{Models using the membrane potential as prediction error (Class 2) }
  \vspace{-0.5cm}
\end{table}

In the second class of spiking predictive coding models that we will review, prediction errors are not encoded by explicitly designated error neurons but instead distributed throughout the network in the membrane potentials of the neurons themselves. In other words, the membrane potentials of the neurons represent the prediction error between the input received by a network and its decoded activity which forms the estimate of (a function of) the input. In this sense, these particular kinds of PC networks unite the encoding and decoding perspectives on neural coding -- they show that efficient encoding cannot be formulated independently from the decoding algorithm itself and that what an efficient code is depends inherently on the decoder used and the task under consideration. Here, we will first show the original formulation of spike coding networks (SCNs). Next, we will discuss their extensions and derivations, as well as other formalisms that calculate prediction errors within the membrane potentials of neurons. 

\paragraph{Spike Coding Networks. }
The derivation of the SCN takes a different perspective than how decoding questions are traditionally formulated. Instead of asking the question 'What is the optimal decoder for this neural activity?', the derivation starts with the question 'Given input signal $s(t)$, a decoder $X$, an error function $E(s, \hat{s})$, and a cost function $C$ that limits the neural activity, what should my network structure look like for the network to optimally (under $E$ and $C$) generate activity that can be decoded to an estimate $\hat{s}(t)$ with decoder $X$?'. In their original formulation, \cite{boerlin2011spike, boerlin2013}, Boerlin, Machens and Denève used this approach, with a linear decoder with exponential decay and an $L2$ loss function, to derive a network where recurrent synaptic connections of neurons that have similar decoding weights are inhibitory, whereas recurrent connections of neurons that have opposite sign decoding weight values excite one another. Each neuron has a local estimate of the network's current estimate $\hat{s}$  through the recurrent input from the other neurons and compares the future estimate given that it would generate a spike or not. Only if the spike would contribute to an improved estimate $\hat{s}$ does the neuron fire a spike. So similarly to the Bayesian neuron (Sec.~\ref{par:BN}), each neuron compares continually the input to its own output and effectively only generates a spike if that contributes to an estimate of the input. So in this case too, as with the Bayesian neuron, the parameter that is being tracked continuously and compared to a threshold, i.e. the `membrane potential', is a prediction error itself. For an explicit derivation, see \textbf{Box 3}. 
This process can also be interpreted as the network tracking the input within a `bounding box' with a precision that is given by the threshold and the number of neurons \cite{calaimGeometryRobustnessSpiking2022}. Note that the computations that the network can perform are not limited to autoencoders (i.e. $s \approx \hat{s}$) but also include polynomial dynamical systems \cite{10_24072_pcjournal_69} as well as optimal control \cite{slijkhuisClosedFormControlSpike2023}. 

Surprisingly, the derived networks described above exhibit properties that were difficult to explain using more traditional decoders. One of these properties is that the networks show irregular activity with a high trial-to-trial variability. Importantly, this irregular activity should not be confused with noise. Rather, it is a result of degenerate coding: since there are multiple `solutions' to the problem of tracking the stimulus, the network will use different solutions given different initial conditions. This over-representation makes the networks also strongly robust to noise, perturbations, and cell loss \cite{barrettOptimalCompensationNeuron2014}. The error-correcting properties of the network also reflect a form of the neurobiological balance of excitation and inhibition \cite{Deneve2016, Schwemmer2014}. Finally, even though the ideal connectivity of the network can be mathematically derived using the method described above, the network connectivity can also be usefully learned via sing local unsupervised learning rules \cite{Bourdoukan2012, alemiLearningArbitraryDynamics2017, brendel2020learning}. 

In their original formulation, SCNs represented a class of networks that can explain observed biophysical properties of biological neuronal networks, such as a balance of excitation and inhibition, large trial-to-trial variability, irregular firing, and robustness against noise and cell loss, in a functional context. This functional context is that of predictive coding, in which each of the neurons tracks the error between (a function of) the input and the network estimate in the membrane potentials of the neurons. These codes are very efficient (in terms of sparsity) since spikes are only fired if they contribute to the network estimate.

\paragraph{Extensions of Spike Coding Networks towards Biological Realism}  
The original formulation of the SCNs showed the fundamental entangling of encoding and decoding, but this was from a biophysical point of view and still rather abstract. For instance, the networks consist of a homogeneous population of leaky integrate-and-fire (LIF) neurons rather than the more complex and heterogeneous populations found in biophysical recordings. Moreover, connections are dense and do not obey Dale's law, i.e. each neuron can be both excitatory and inhibitory. Several authors have extended the SCN framework to investigate the feasibility of this framework to more biologically realistic settings. 

With respect to including more biologically realistic neuron models, several extensions of the SCN framework have been made. Zeldenrust et al. \cite{zeldenrust2021} used `filter-and-fire' models with different kernels (similar to the matching pursuit formulation in auditory signals as in Sec.~\ref{sec:Historical} \cite{Smith2005}) rather than LIF models and found that increasing the heterogeneity of the neuron population increased the network's efficiency and robustness against noise. Schwemmer et al. \cite{Schwemmer2014} developed an implementation with conductance-based models, which functions so long as the kinetics of synapses are faster than those of the signal. Gutierrez and Denève \cite{gutierrezPopulationAdaptationEfficient2019} showed that spike-frequency adaptation, a widespread property of biological neurons \cite{Gutkin2014}, can be understood as a solution to the cost-accuracy trade-off. 

The connectivity of the original SCN models was dense and needed instantaneous connections to prevent a 'ping-pong' effect, in which neural populations that code for opposite features in the input spike in alternating time bins, resulting in inaccurate estimates and very high firing rates. Koren and Denève showed that synaptic delays could be included in SCNs \cite{koren2017computational} so long as a spike cost is included in the objective function, limiting firing rates similarly to the spike-frequency adaptation discussed above. Alternatively, Chalk et al. \cite{Chalk2016} implemented an SCN with synaptic delays that complied with Dale's law. In these networks, gamma oscillations occur as a hallmark of an efficient code. Similarly, Koren and Panzeri \cite{koren2022biologically} defined separate loss functions for excitatory and inhibitory populations with short synaptic decay time constants, deriving a network that can perform general linear transformations that complies with Dale's law and uses a generalized LIF model. Buxó and Pillow \cite{rullanbuxoPoissonBalancedSpiking2020} implemented another solution to the fast recurrent connectivity problem; instead of a deterministic spike firing rule, they implemented conditional Poisson firing, i.e. soft instead of hard thresholds, which allowed for delays in the synapses without the `ping-pong' effect occurring. Finally, Timcheck et al. \cite{timcheck2022optimal} generalized the problem of instantaneous connections and derived an analytical expression for how coding fidelity depends on noise levels and delays in SCNs; this work demonstrated that there is an optimal noise level given the synaptic delay between neurons. 

In conclusion, the rather abstract and simplified original formulation of SCNs can be made more biologically realistic by including more realistic neuronal models and connectivity structures. However, whether the brain actually employs forms/structures much like SCNs, as well as how this can be experimentally assessed, remains an object for further study and investigation.

\begin{tcolorbox}[width=\linewidth, sharp corners=all,left=0pt,right=0pt,boxrule=0pt, colback=white!95!black]
\textbf{Box 3: Derivation of the SCN.} Although choices for decoders, error functions, cost functions, etc. change according implementation, an SCN's derivation adheres to the 4 steps below; illustrated via the example of \cite{boerlin2011spike, boerlin2013}.
\begin{itemize}
    \item{\textbf{Step 1: Define a decoder, error function $E$, and cost function $C$.}} In this case, we define a linear decoder with exponential decay. The network consisting of $N$ neurons $i$ generates output spike trains $o_i(t) = \sum_\textrm{spikes k in neuron i} \delta(t-t^k_i)$. This is converted to a firing rate by a leaky integration: 
    \begin{equation}\label{eqn:scn_rate} \frac{dr_i}{dt} = \frac{-r_i}{\tau} + o_i(t),  \end{equation}  
    where $\tau$ is the time constant; 
    or similarly
    \[ r_i(t) = o_i(t) \circledast h(t) =  \int_0^t h(t') o_i (t-t ')  d t', \] where $\circledast$ denotes a convolution and $h(t)$ an exponential kernel with time constant $\tau$. The firing rates of each neuron of the network is decoded using a linear decoder: 
    \[ \hat{s}(t) = \sum_{i=1}^N w_i r_i (t).\] We assume the network is an autoencoder, whose task it is to encode the input $s(t)$ by reducing the error $E$, which is determined using the $L2$ norm:
    \begin{equation} \label{eqn:scnerror}
         E(t) =(s(t) - \hat{s}(t))^2 . \end{equation}
    Finally, we assume a simple constant cost $C$ for each spike, that for now does not depend on (the history of) the firing rates (but see for instance \cite{gutierrezPopulationAdaptationEfficient2019} for other choices). 
\item{\textbf{Step 2: Determine the contribution of a single spike to the error.}} If neuron $i$ fires at time $t$, what does this contribute to the error? The error without this extra spike is given by equation \ref{eqn:scnerror}. The error with this extra spike is given by
\[E(t)^\textrm{spike at t} = (s(t) - \hat{s}(t) - w_i h(t))^2  = (s(t) - \hat{s}(t) - w_i)^2,\]
where in the right hand side the exponential decay can be ignored because the error is considered at time $t$ (but see for instance \cite{zeldenrust2021} for a longer integration window).
\item{\textbf{Step 3: Efficiency, every spike should reduce the error.}} Neuron $i$ should only spike at time $t$ if
\begin{equation} \label{eqn:spike_ineq}
\begin{split}
    E^\textrm{no spike neuron $i$ at $t$} &> E^\textrm{spike neuron $i$ at $t$} + C \\ (s(t) - \hat{s}(t))^2  &> (s(t) - \hat{s}(t) - w_i )^2  +C \\
0 &> \left( - 2 w_i s(t) + 2 w_i \sum_{j=1}^N w_j r_j(t) +w_i^2 \right) +C \\  w_i s(t) - w_i \sum_{j=1}^N w_j r_j(t)  &= w_i (s(t) -\hat{s}(t) ) >  \frac{w_i^2 +C}{2}\end{split}   \end{equation} 
\item{\textbf{Step 4: Derive 
the membrane potential, threshold, recurrent connectivity, network dynamics}} Typically, the threshold $\Theta_i$ of the neuron is  defined as the right-hand side of inequality (\ref{eqn:spike_ineq}), which is time-independent. The time-dependent left-hand side is called the `membrane potential' $V_i$ and is also the prediction error. Thus, the membrane potential is an integration of  incoming input ($w_i s(t)$) compared with a network estimate ($w_i \sum_{j=1}^N w_j r_j (t)$). To obtain network dynamics, we  take the time-derivative:
\[\begin{split} \tau \frac{dV_i}{dt} & =\tau w_i \frac{d s}{dt} - \tau w_i \sum_{j=1}^N w_j \frac{d r_j}{dt} \\
&= \tau w_i \frac{d s}{dt} + w_i \sum_{j=1}^N w_j r_j (t) - \tau w_i \sum_{j=1}^N w_j o_j (t) + w_i s(t) - w_is(t) \\
&= -V_i (t) + I(t) -\tau w_i \sum_{j=1}^N w_j o_j(t), \end{split}\]
where we used the definitions of firing rate (equation (\ref{eqn:scn_rate})) and membrane potential (equation (\ref{eqn:spike_ineq})) and define feed-forward input current as $I(t) =  w_i (\tau\frac{ds}{dt} + s(t))$. Note that the neuron dynamics and network connectivity are completely determined by the choice of decoder, cost function, and error function. 
\end{itemize}
The extensions and variations of SCNs discussed in Sec.~\ref{sec:Class2} all follow these steps. They define different error functions, cost functions or decoders, but they all follow the same basic derivation.
\end{tcolorbox}

\paragraph{Dendritic Computation of Prediction Error}
Next to the SCNs discussed above, several authors have proposed that dendrites are essential in computing prediction errors at the neuron level. Thalmeier et al. \cite{thalmeierLearningUniversalComputations2016} generalized SCNs to universal computations (including non-linear dynamical systems) that are linearly decoded. As a result of the linear decoding, the non-linear transformations must be computed in the network, which was, in this case, done by either saturating or non-linear dendrites crafted at the neuron level.  Mikulasch et al. \cite{mikulaschLocalDendriticBalance2021, mikulasch2023} did not use a single-layered network to predict a (function) of an input but instead derived a classical hierarchical predictive coding network, where each layer sent a feedback prediction signal of its input to the previous layer and a feed-forward error signal to the next layer \cite{rao1999predictive}. Instead of using separate populations of `error neurons' and `prediction neurons' in each layer, in this framework, predictions were computed in the somas and errors were computed in apical (feed-forward errors) and basal (feedback errors) dendrites of neurons as the difference between the input (error) and the somatic prediction. Recently, Zhang and Bohte \cite{zhang2024energy} took an (bi-objective) optimisation approach: they trained a multi-layered network with multi-compartment neurons to minimize energy consumption 
in tandem with task-relevant (reconstruction) objective 
using gradient optimisation. In these networks, feed-forward signals were sent to the somas while feedback signals were sent to the dendrites. After training, these networks exhibited two hallmarks of predictive coding: they made internal predictions of expected stimuli and elicited different responses to expected and unexpected stimuli. So in these neural architectures, predictive coding-like behaviour is a result of energy optimization (rather than the other way around). In essence, these three implementations of predictive coding show that the addition of dendritic non-linearities can greatly increase the computational repertoire of predictive coding networks, especially when computations are performed hierarchically in consecutive layers.

\paragraph{Conclusion} The derivation of `class 2' predictive coding networks all follow a similar path: a goal computation, an error function, a cost function and decoder are defined, and the neuron and network properties are derived or learned as a result of this. In the resultant network structures, neurons keep track of the prediction error in their membrane potential and only spike if this reduces the error, making these networks inherently very efficient. The connectivity and properties of the network depend on the choice of computation, error, cost and decoder, but the networks share a tight balance between excitation and inhibition, irregular firing, a large trial to trial variability and robustness against noise and neuron loss.

\section{Implicit Prediction Error Encoding (Class 3)}
\label{sec:Class3}

\begin{table}[!t]
  \label{tab:modelsimplicit}
  \footnotesize
  \centering
  \resizebox{\columnwidth}{!}{%
  \begin{tabular}{cccccc}
	\toprule
	Authors  & Prediction & Neuron type & Dale’s law?  & Learning rule & Explained phenomena/Application\\
	\midrule
	   \thead{Sinapayen et al. \cite{sinapayen2015learning, sinapayen2017learning} \\Masumori et al. \cite{masumori2019}} & Lateral input & Izhikevich & Yes & STDP & \thead{How to predict temporal sequences \\and suppress them}\\
    \midrule
     \thead{Schulz et al. \cite{schulz2021generation}\\ Zhu and Rosenbaum \cite{zhu2022evaluating}} & Lateral and top-down inputs & Exponential IF & Yes & STDP & \multirow{3}{*}{\thead{How to generate \\prediction error-like \\responses}}\\
    \cmidrule{1-5}
    Asabuki et al. \cite{asabuki2023learning} & Lateral input & Conditional poisson neuron & Yes & Custom rule from \cite{asabuki2023embedding} & \\
    \cmidrule{1-5}
    Van driel et al. \cite{van2023prediction} & Lateral input & Conditional bayesian-based neuron & No & Fixed weights & \\
    \midrule
    N'dri et al. \cite{ndri2023} & Lateral and top-down inputs & LIF & No & STDP & \thead{How to learn V1-like responses \\that exhibit surround suppression}\\
    \midrule
    Saponati and Vinck \cite{saponati2023sequence} & Feedforward input & LIF & Yes & Discrete local predictive rule & \thead{How to predict temporal sequences\\ and learn to anticipate inputs}\\
    
	\bottomrule 
  \end{tabular}
  }
  \vspace{0.1cm}
  \caption{Models using an implicit prediction error encoding (Class 3)}
\end{table}

In the predictive coding models reviewed so far, errors were always explicitly encoded in some fashion. Class 1 used specific error units whereas class 2 derived spiking networks that encoded errors within neuronal membrane potentials. While these models for error encoding are attractive, they are not the only ones proposed in the literature. Due to the link between predictive coding and the general efficient coding hypothesis, we see the influence of previous spiking models of efficient sensory coding (see Sec.~\ref{sec:Historical}) on recent efforts that fall under a third class of models that propose to implement spiking predictive coding with different error encodings. Most of these models exhibit prediction error-like responses with no explicit definition of error encoding. Instead, they implicitly encode errors; for instance, this is done via competition between plastic excitatory and inhibitory synapses. In addition to this implicit encoding via excitation/inhibition, alternative methods have been proposed including the implementation of correction neurons or the design of a predictive learning rule. Methods of this third class are amongst the most recent predictive coding approaches (see timeline in Fig.~\ref{fig:timeline}).



    
    
\paragraph{Efficient Coding through Suppression.}
According to the efficient coding hypothesis, an efficient code would represent a stimulus with as few spikes as possible. In the context of predictive coding, this would be achieved with a direct suppression of redundant codes through some sort of prediction neurons. Sinapayen et al. proposed ``learning by stimulus avoidance'' (LSA) \cite{sinapayen2015learning, sinapayen2017learning}, an STDP-based spiking neural network learning framework that exhibited predictive coding-like effects such as suppression. LSA is based on the idea that neurons learn to respond to external stimulation by having neural responses that minimize the chance of receiving more excitation. This learning framework was derived from the experiments of Shahaf and Maron on cortical neurons of rats \cite{shahaf2001learning, marom2002development}. In those experiments, neurons were excited until the rats performed a target behavior. As soon as the target behavior was achieved, stimulation was removed. Hence, neurons learnt to optimally respond to a stimulus by aiming to remove it. During the training of actions, stimuli were therefore removed only when the expected behavior was achieved by the overall network (of Izhikevich neurons). At the neuron scale, STDP naturally leads to stimulus removal as inputs that lead to a desired action have their synaptic strength increased. This work demonstrated that this effect can also appear at the scale of a network and that populations can essentially learn selective behaviors. 
While LSA was introduced as a control approach without an explicit predictive aspect, Masumori et al. showed that LSA, through the inclusion of inhibitory STDP, could be used to perform predictive coding \cite{masumori2019}. They connected a pool of input excitatory neurons to inhibitory neurons. After presentations of a pair of signal/target stimuli, inhibitory neurons learned to predict (and therefore suppress) target signals. The authors also observed that, if synaptic delays were utilized, their network could suppress signals relatively far in time. 

Recently, N'dri et al.\ proposed ``predictive coding light'' (PCL) \cite{ndri2023}. PCL questions the idea of only passing prediction errors to higher processing stages. Instead, a compressed representation of the actual perceptual content was passed to higher processing levels. The compression is achieved through the interaction of lateral and top-down inhibitory connections, which learn to suppress the most predictable spikes via a form of inhibitory STDP. The idea can be effectively distilled to the notion that highly predictable spikes carry little additional information and should therefore be suppressed to save precious energy (expended during neural computations). In this sense, PCL is deeply rooted in efficient coding ideas \cite{Barlow1961} as well as sparse coding approaches \cite{olshausen}.
Nevertheless, PCL does not attempt to be a faithful model of biology. For example, in the implementation of N'dri et al., there is no distinction between excitatory and inhibitory neurons, but there exists only a single neuron type, which is modeled as an LIF neuron. These neurons excite neurons in higher processing stages via their feedforward connections and inhibit neurons at the same or lower processing stages, thus violating Dale's law. Despite this modeling limitation, an attractive feature of PCL is that all connections are learned with biologically plausible local STDP rules.

Notably, the authors of PCL demonstrate that networks within this framework manage to roughly halve the number of spikes used to represent visual input from a neuromorphic vision sensor (i.e., a dynamic vision sensor) with only a moderate loss of information as empirically assessed under different classification tasks.
Interestingly, the PCL network of early visual processing also learns biologically plausible simple and complex cell receptive fields and reproduces surround suppression and orientation-tuned suppression phenomena, which are considered to be likely signatures of the brain employing a predictive coding approach.

\paragraph{Prediction Errors through Inhibitory Plasticity.} It has been observed experimentally that neuronal responses change as a function of the frequency at which a stimulus is presented. Neurons repeatedly excited with a certain stimulus exhibit a decreased response (``repetition suppression'') compared to those excited with a novel or deviant stimulus. This change in response signals a decreasing prediction error in the predictive coding framework. While this is generally explained with an increase of the response of error neurons, Schulz et al. challenged this idea by proposing inhibitory plasticity as an explanation of such cortical prediction error responses \cite{schulz2021generation}. 
The authors specifically modeled a biologically plausible spiking network model of the mammalian cortex in which neurons were recurrently connected via both excitatory and inhibitory pathways. 
The different synaptic connections in this model were plastic and learnt via a triplet spike timing-dependent plasticity rule (triplet STDP) \cite{gjorgjieva2011triplet}. 
Schulz et al.\ notably showed, in a series of experiments, that not only the learnt inhibition generated prediction errors/novelty responses but also explained stimulus-specific adaptation (SSA). 

Zhu and Rosenbaum \cite{zhu2022evaluating} also analyzed the emergence of such predictive coding-like effects with homeostatic plasticity. Similar studies were also conducted using rate-coded neurons 
\cite{hertag2020learning, hertag2022prediction}. 
While the authors considered populations of neurons that received fixed bottom-up input and top-down feedback, the computation of prediction errors did not arise from a direct mismatch of the signals but rather from the learning of recurrent inhibition via STDP. In their experiments, the authors showed that, when trained with pairs of bottom-up input and top-down feedback, inhibition reduced activity to target firing rates. However, they further observed that when bottom-up and top-down signals no longer matched, inhibition failed to stabilize network activity, which led to the emergence of prediction errors. In addition, these researchers analyzed their model's network dynamics by deriving rate codes from their spiking model using mean-field theory. 

Similarly, Asabuki et al. also studied the emergence of prediction errors but particularly in networks that received no top-down feedback \cite{asabuki2023learning}. 
While in the classical predictive coding view, prediction errors arise from the arrival of feedback from deeper brain areas, recent experimental results intriguingly indicate that top-down feedback might not be necessary for their emergence. A circuit of excitatory and inhibitory neurons recurrently connected might be enough to observe prediction error responses. The authors therefore considered a recurrent circuit of a pool of excitatory neurons that received bottom-up excitation and could recurrently excite each other. The researchers excited a pool of inhibitory neurons that were also connected with each other. At each point in time, the membrane potential of a neuron depended only on its immediate excitatory and inhibitory inputs. The authors' model notably spiked under an inhomogeneous Poisson process, where the firing rate of a neuron depended on the value of its membrane potential. They developed a custom learning rule \cite{asabuki2023embedding} in which recurrent synaptic weights were potentiated if they could predict the activity of postsynaptic neurons. The authors showed that such a framework facilitated the learning of synaptic weight values that resulted in the emergence of a wide variety (biphasic, expectation dependent, context-dependent) of prediction error signals. The authors explained that prediction errors in their framework came from changes in the balance of excitatory-inhibitory currents under conditions different from training. 

\paragraph{Predictions and Prediction Errors through Learning Rule and Correction Neurons.} 
Taking inspiration from general predictive processing, Saponati and Vinck proposed a predictive learning rule. The network in this work adapted its synaptic weight values by anticipating future inputs by firing ahead of time \cite{saponati2023sequence}. The objective of any single neuron in this model was to predict its next inputs from its membrane potential and its feedforward synaptic weights such that it minimized 
prediction errors directly encoded in the loss function of the network. The derived learning rule potentiated connection strengths from inputs that were found to be useful in making predictions of future inputs, whereas inputs that did not enable the prediction future inputs had their connections depressed. The authors demonstrated that, in a recurrently-connected spiking neural network of LIF-like neurons, such a predictive learning rule permitted the forecasting/prediction of future input sequences while also accounting for the emergence of spike timing-dependent plasticity rule patterns. 

Van Driel et al. alternatively proposed an explanation of prediction mismatch responses/prediction errors with ``correction neurons'' \cite{van2023prediction}.
They considered a pool of recurrently-connected spiking LIF-like neurons that received feedforward excitation as well excitatory and inhibitory feedback (predictions). All connections in this model were fixed. Predictions or feedback lead to early (anticipatory) firing of neurons and, when these predictions and the input did not match, a class of neurons called ‘correction neurons’ were activated. Those neurons bore similarity to error neurons but, instead of signaling an error, they fired to signal that anticipated spikes were ``wrong''. Those spikes have thus a functional use as a correction of the neural code instead of being a coding error. Van Driel et al. showed that such a mechanistic network could account for prediction errors in a minimal network; this work desirably reproduced the prediction errors/prediction mismatch responses' experiment on a mouse’s visual cortex in Fiser et al. \cite{fiser2016experience}.

\paragraph{Conclusion.}
The methods proposed in this third class of models questioned the need for explicit error encoding in general. For instance, without any explicit definition of an error, STDP-learnt inhibition alone is able to learn efficient sensory codes in which redundant stimuli are suppressed and only unpredictable stimuli generate spikes. Similarly, when looking at the response of neurons to frequent and deviant stimuli, methods of class 3 showed prediction mismatch responses -- that we defined in this work as prediction errors -- that arose from the competition of, again, STDP-learnt excitatory and inhibitory synapses. The different uses of STDP for predictions are further reminiscent of Rao and Sejnowski's earlier works on predictive learning \cite{rao1999predictivespike,rao2001spike,rao2002}(see Sec.~\ref{sec:Historical}). We also reviewed under class 3 various novel approaches that learnt predictions through a predictive rule, which further replicated observations of prediction errors through `correction neurons' instead of classical `error neurons'. In general, the results achieved by methods in this third class suggest that an explicit error encoding might not be a vital requirement for predictive coding with spikes.

\section{Neuromorphic Applications of Spiking Predictive Coding}
\label{sec:Applications}


When performing predictive coding, predictions and error computations are performed locally by neurons that explicitly connect to one another. This has even been argued to offer an alternative to backpropagation, the workhorse of deep learning  \cite{millidge2020relaxing,alonso2021tightening,zahid2023predictive,salvatori2023brain,ororbia2024review}. The local computations performed by predictive coding makes it not only a biologically plausible alternative to backpropagation but also a fitting candidate for applications on neuromorphic hardware. Despite these attributes, however, predictive coding has been only scarcely implemented on neuromorphic platforms. It is well known that standard deep learning approaches suffer from the weight transport problem as well as expensive computational costs for training (such as the forward and backward locking problems \cite{salvatori2023brain,ororbia2024review}), which render their implementation on neuromorphic hardware complicated. Very recent efforts to instantiate predictive coding neuromorphically have also begun to call into question if predictive coding itself also still suffers constraints similar to those that hinder backprop-based deep networks \cite{flugel2023feed,hagiwara2024design}; these efforts generally offer workaround solutions to these potential constraints, using schemes as feedback alignment and neuronal reservoirs to implement a more hardware-friendly version of predictive coding. 
We review here such works as well as applications of neuromorphic predictive coding with a mapping of predictive coding algorithms based on Friston's work to spiking neurons, including some implementations on hardware. 

\paragraph{Predictive Coding under Hardware Constraints.} Hagiwara et al.\ evaluated the performance of classical predictive coding under hardware constraints \cite{hagiwara2024design}. To reduce the training computational costs of deep neuronal models on neuromorphic platforms, techniques such as feedback alignment and reservoirs are conventionally used. Feedback alignment denotes the replacement of the backpropagated errors by randomly initialized, fixed synaptic weights. Temporal reservoirs, on the other hand, denote randomly initialized networks that are used during feedforward propagation to better represent sequential information. Due to the non-adaptive, random nature of their synaptic weights, feedback alignment and reservoirs result in little computational cost during training. Hagiwara et al.\ \cite{hagiwara2024design} proposed applying such engineering techniques to predictive coding in an effort to construct a network that is inexpensive during training. They validated their work on image classification tasks and in tasks involving the prediction of dynamic signals, and showed that such types of predictive coding networks yield similar results to classical predictive coding. Inspired by the prominent results of reservoirs in information processing, Ishikawa et al.\ also combined reservoir computing with predictive coding \cite{ishikawa2024integrating}. They recently proposed an implementation of predictive coding in which prediction error units updated a dynamical reservoir of spiking neurons. This reservoir would then generate predictions of the input.

\paragraph{Implementations of Spiking Predictive Coding on Hardware.} In Friston's formulation of predictive coding, a network operating under the free energy principle minimizes the so-called variational free energy which calculates the difference between the priors (an internal model of the causes behind an observation) and the posterior (how the observation actually evolves as inferred by data/evidence). While this Fristonian view of predictive coding has been generally applied to rate coding modeling approaches, Palacios et al.\ extended it to networks of spiking neurons \cite{palacios2019emergence}, similar to the work reviewed under `Class 1' models. 
In their framework, each neuron tries to predict when other neurons will spike by minimizing the variational free energy via so-called ‘variational message passing’. In this approach, a network knows the exact generative model from which an observation 
is generated,
as well as the inverse (inference/recognition) model. The network makes use of this inverse model: each time a new observation 
is received, the network infers the hidden causes (spikes) behind it. It uses this information, the spiking rate of the network, and predictions of future activity 
to update the firing rate of all neurons. Neurons then spike only when their firing rate is above a particular threshold. 
Such a scheme has been shown to lead to synchronous networks that are robust to perturbations and develop sparse connectivity between neurons. Gandolfi et al.\ built on this work by reproducing so-called delayed eye-blink classical conditioning effects with a cerebellum-like network structure of mutually inferring free-energy-minimizing neurons \cite{gandolfi2022emergence}. Their network consisted of 
neurons with both excitatory and inhibitory connectivities that respect Dale’s law. Crucially, Gandolfi et al.\ implemented digital neurons on low-power microcontrollers based on these computational principles. This neuromorphic hardware instantiation of predictive coding strongly reduced the number of neurons required (and therefore also power consumption) compared to other methods that replicated the eye-blink classical conditioning effects. The results obtained on hardware furthermore outperformed software results. Critically, these results mark an important step towards low energy consumption and increased performance of brain-inspired efficient computing on neuromorphic hardware.

With respect to other neuromorphic applications of predictive coding, Feng et al.\ proposed a spiking model of robot pain inspired by the free energy principle \cite{Feng2022pain}. In their brain-inspired robot pain spiking neural network (BRP-SNN), `pain' resulted from the mismatch between sensory inputs and their prediction due to physical injury. 
The model was implemented as a feedforward SNN with LIF neurons and population coding. Prediction neurons adapted via STDP to predict sensory data in a normal robot state. A population of error neurons received inputs from both prediction neurons and sensory neurons through excitatory and inhibitory connections, respectively.
Error neurons fired in the case of prediction errors due to robot injury and triggered the pain module through excitatory connections.
Notably, this model responded to both actual injury and potential injury as detected by perceptive cues associated with previous injuries through STDP learning. The proposed model was validated on a humanoid robot equipped with proprioception and visual cues.


\paragraph{Conclusion.} 
In this final section, we reviewed efforts aimed at understanding how predictive coding would apply when considering hardware constraints such as training computational resources as well as other efforts that took inspiration from the free energy principle to propose spiking implementations in both software and hardware. The scarcity of hardware implementations of spiking predictive coding suggests an important open area of research for fast, efficient and biologically plausible computing applied to real-world tasks.


\begin{figure}
\centering
\includegraphics[width=15cm]{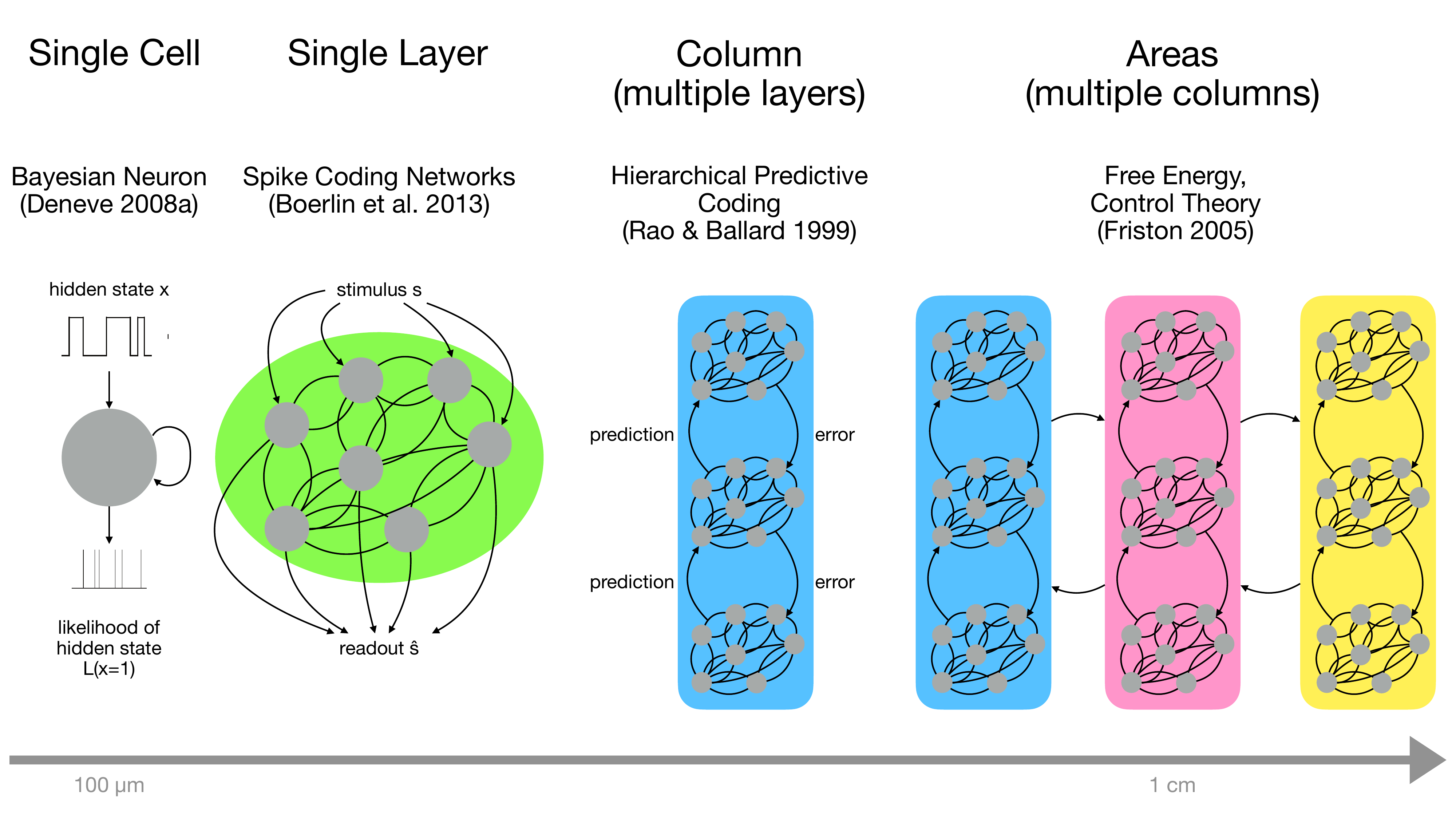}
\caption{{\bf Cognitive and spatial organization of spiking predictive coding models.} Although we organize the models discussed in this survey according to how the prediction error is formulated/represented, alternative organizations are possible. For instance, the models presented here can be organized at different spatial and cognitive levels/resolutions. 
 }
\label{fig:cognitive}
\vspace{-0.5cm}
\end{figure}

\section{Conclusion and Outlook}
\label{sec:Conclusion} 


In this article, we have surveyed approaches that have aimed to implement predictive coding (PC) in terms of spiking neural networks. There are two main reasons for the growing interest in this topic. First, predictive coding ideas feature prominently in theories of brain function and it is of great importance to understand their possible implementations at the level of spiking neurons in the brain; this is a computational neuroscience motivation. Second, there is enormous, growing interest in constructing more energy-efficient AI systems and spiking neural networks are interesting potential candidates to facilitate this, as reflected in the growing interest in neuromorphic hardware; this is a neuromorphic engineering motivation. Given these two distinct motivations, it is not surprising that the literature on the topic is somewhat scattered across different fields and venues. This article surveys this literature within a single unifying framework based on predictive coding principles.

We chose to organize this survey based on how prediction errors are encoded under various computational modeling approaches and schemes. Specifically, we decomposed these under three broad categories: models utilizing specific error units (Class 1), models encoding errors through a unit's membrane potential (Class 2), and implicit error models without any direct error encoding mechanisms (Class 3). This was a pragmatic choice meant to highlight the fundamental differences between spiking PC approaches. However, other organizations are also been possible. For instance, we found a dichotomy between approaches that aim to better understand implementations of PC in the brain through computational modeling (thus reflecting  computational neuroscience  motivations) and approaches that focus on technological advances (thus reflecting neuromorphic engineering motivations). In the present organization of the survey, these approaches have been interspersed among the classes, which could be considered a shortcoming. In fact, instances of these two approaches are somewhat difficult to compare since they use very different ``measures'' of success. While computational neuroscience puts an emphasis on improving our understanding of biological phenomena (reproduction and explanation of biological data, generation of testable predictions), what matters for neuromorphic engineering is performance in applications (accuracy, speed, energy efficiency). Nevertheless, we view the organization presented in this review as an important starting point and stepping stone for fostering cross-fertilization across these two perspectives. Ultimately, computational neuroscience research could greatly benefit from neuromorphic computing hardware to speed up simulation times, which are often a limiting factor in computational neuroscience research and scientific inquiry. At the same time, neuromorphic engineering may harvest novel ideas from computational neuroscience as to what biological mechanisms may be worth implementing in the next generation of neuromorphic systems and hardware platforms.

Our division of modeling approaches into the three different classes raises the obvious question of whether one class should be considered superior to the others; what is the best modeling paradigm for spiking predictive coding? At this point, we do not think that any approach has sufficiently demonstrated its superiority to rule out the other ones, particularly, as mentioned before, approaches across these classes use different measurements of success depending on their scientific goals and the research questions that they seek to answer. Furthermore, it is important to realize that the answer may depend on which of the two motivations -- neuroscience versus neuromorphic engineering -- is driving a specific research effort. It may well turn out that one class of approaches turns out to be a better model of PC in the brain whereas a different class prevails in technical applications in neuromorphic systems such as neurorobots. After all, recalling David Marr's classic analogy, we do not build airplanes with feathers and flapping wings. 

Even within the field of computational neuroscience, the different implementations of spike-based PC do not all offer explanations at the same level of abstraction and at the same spatial scale (see Fig.~\ref{fig:cognitive}). Whereas the Bayesian neuron and spike coding networks focus on how errors are represented in neurons and how single functions of input stimuli are represented, hierarchical predictive coding models offer explanations at higher cognitive and network levels of abstraction and bigger spatial scales, specifically shedding light on how information is shared between layers and across areas. Finally, many control theory and free energy implementations of predictive coding operate on even higher levels of abstraction (as well as at an even larger spatial scale), incorporating actions and closed-loop interactions with the environment (e.g., active inference \cite{friston_free-energy_2010,Sajid_active_inf_2021} and active predictive coding \cite{Rao-APC-2024,Rao-NN-2024}). 

Related to the above points, different hard- or wet-ware components {\em invite} but also {\em discourage} certain solutions to a given problem. For instance, a particular PC formulation may be computationally efficient and work well in practice on certain spiking neural network hardware platforms yet might be impossible to realize in biological nervous systems with all their constraints on, e.g., connectivity, time scales of signal transmission and processing, etc. Ignoring such constraints can quickly lead to research `dead ends`, especially if the goal is to understand how PC is implemented in the brain. Conversely, adhering to constraints that exist in biology but that could be overcome in engineered computing hardware runs the risk of overlooking particularly efficient and powerful solutions to certain engineering problems. At the same time, neuromorphic hardware may have its own constraints stemming from, e.g., the chip manufacturing process, that have no correspondence in biology. Despite these differences, we feel that the two perspectives still have much to offer to each other for the foreseeable future as many fundamental questions remain unanswered and possibly useful transdisciplinary lines of scientific inquiry could emerge in the future that result in new goals and questions, such as those related to neuromorphic constructions of PC in biotic or chimeric substrates \cite{ororbia2023mortal} .

What the computational and neuromorphic perspectives do share, however, is the fundamental conclusion that energy efficiency on the one hand, and  task formulation and implementation on the other (or formulated differently, encoding and decoding) should not be considered independently of one another. Predictive coding and its preference for suppressing predictable stimuli generally results in the decorrelation of network activity \cite{chalkUnifiedTheoryEfficient2017}. This can increase efficiency for the representation of stimuli but not for working memory-like tasks, as has been shown in the case of neural heterogeneity \cite{zeldenrust2021, gastNeuralHeterogeneityControls2024}. Spike coding networks in particular show explicitly how entangled the encoding and decoding problems are. Therefore, we conclude that what constitutes an efficient code cannot be viewed independently from task definition and implementation.

Finally, given that the research topic of spiking and neuromorphic predictive coding (and predictive processing) is still relatively young, as indicated by the overall relative dearth of explicit efforts in this direction, an important underlying aim of this survey is to provide a unifying point-of-view of what has been done to encourage and inspire the upcoming generations of scholars, scientists, and engineers to work on crafting spiking predictive coding models and their physical instantations. Looking forward, making new advances and developments in, as well as answering key questions related to, spiking predictive coding could prove particularly invaluable for the field of brain-inspired computing \cite{mehonic2022brain}, biomimetic general intelligence and bionic engineering \cite{ororbia2023mortal}, and green artificial intelligence \cite{schwartz2020green}.


\bibliographystyle{acm}
\bibliography{ref}

\newpage
\input{appendix}

\end{document}
